\documentclass[12pt,a4paper]{article}
\usepackage{amsmath,amsfonts,amssymb,amsthm}

\newtheorem{theorem}{Theorem}[section]
\newtheorem{remark}[theorem]{Remark}

\newtheorem{lemma}[theorem]{Lemma}

\newcommand{\R}{{\mathbb R}}
\newcommand{\N}{{\mathbb N}}

\newcommand{\C}{{\mathbb C}}
\newcommand{\x}{{\bf x}}
\newcommand{\cst}{{\rm const}}
\newcommand{\A}{{\bf a}}

\newcommand{\y}{{\bf y}}
\newcommand{\z}{{\bf z}}

\newcommand{\tr}{{\rm Tr}}
\newcommand{\fl}{{\rm fl}}
\newcommand{\Fl}{{\rm Fl}}
\begin{document}

\noindent  \centerline {\textbf{\Large Generalized susceptibilities for a
perfect
\break quantum gas}\\}
\bigskip

\qquad \quad\qquad \qquad\qquad December the 9-th, 2003

\vspace{0.5cm}

\noindent \textbf{Philippe Briet
\footnote{PHYMAT-Universit\'e de Toulon et du Var, Centre de Physique
Th\'eorique-CNRS and FRUMAM,
Campus de Luminy, Case 907 13288 Marseille cedex 9, France; e-mail:briet@univ-tln.fr},
    Horia D. Cornean\footnote{Institut for Matematiske Fag,
    Aalborg
    Universitet, Fredrik Bajers Vej 7G, 9220 Aalborg, Danmark; H.C. is
    partially supported
     by  MaPhySto -- A Network in Mathematical Physics and
     Stochastics, funded by The Danish National Research Foundation; e-mail:
    cornean@math.auc.dk}, and Delphine Louis\footnote{Universit\'e de Toulon
et du Var,
PHYMAT-Centre de Physique Th\'eorique-CNRS and FRUMAM, Campus de Luminy, Case 907
                               13288 Marseille cedex 9, France;
                               e-mail:louis@cpt.univ-mrs.fr}}

\vspace{0.5cm}

%%%%%%%%%%%%%%%janvier.2004%%%%%mprf%%%%%%%%%%%%%%%%%%%%%%%%%%%%%%%%%%%%%%%%%%
\vskip3mm\noindent {\bf Abstract: }%
%%%%%%%%%%%%%%%%%%%%%%%%%%%%%%%%%%%
The system we consider here is a charged fermions gas in the effective
mass approximation, and in grand-canonical conditions. We assume that
the particles are confined in a three dimensional cubic box $\Lambda$
with side $L\geq 1$, and subjected to a constant magnetic field  of
intensity $ B \geq 0 $.
Define the grand canonical generalized susceptibilities $\chi_L^N$,
$N\geq 1$, as
successive partial derivatives with
respect to $B$ of the grand canonical pressure $P_L$. Denote by
$P_{\infty}$ the thermodynamic limit of $P_L$. Our main result is that
$\chi_L^N$ admit as thermodynamic limit the corresponding partial
derivatives with respect to $B$ of $P_{\infty}$. In this paper we only give
the main steps of the proofs, technical details will be given elsewhere.

\vspace{0.35cm}


\vskip3mm\noindent \section{Introduction and results}%
%%%%%%%%%%%%%%%%%%%%%%%%%%%%%%%%%%%%%%%%
In this paper, we are interested in the
thermodynamic behavior of perfect Fermi  gas in the presence of a constant
magnetic field ${\bf B}$ at temperature $T>0$ and chemical potential $\mu$
fixed. Although the particles have electric charge so that they can
interact with the external magnetic field, we neglect all
self-interactions and work in the effective mass approximation. We
also neglect the spin, since it does not change the nature of our
results.

Consider that the gas is confined in a three dimensional cubic box
$\Lambda$ of side $L\geq 1$, centered at  the origin. The constant magnetic
field is ${\bf B}:=B {\bf e}_3$ where ${\bf e}_3:=(0,0,1)$ is the third
vector
of the canonical base of $\R^3$. Denote with $c$ the speed of light,
$e$ the electric charge which is supposed the same for each particle,
and define the Larmor frequency $\omega:=( \frac{e} c)B\geq 0$. We
associate to ${\bf B}$ the  magnetic vector potential $\A$ defined by:
\begin{equation} \label{potm}
B\A(\x):=\frac{B}{2}\,{\bf e}_3\wedge \x.
\end{equation}
The operator $H(\omega):=\frac{1}{2}\,(-i\nabla-\omega \A)^2$ is
essentially self-adjoint on $C_0^{\infty}(\R^3)$ [K]. Denote by $H_L(\omega)$
the
restriction of this operator to $C_0^{\infty}(\Lambda)$. The
Hamiltonian of our one-particle-problem is the self-adjoint
extension of $H_L(\omega)$ with Dirichlet boundary conditions. We will
use the same notation  for $H_L(\omega)$ and  for  its self-adjoint
extension.

Let $z=e^{\beta\mu}$  be the
fugacity (here $\beta=1/T>0$). We will allow $z$ to take complex
values, i.e. $z\in
D:=\C\backslash\,]-\infty,-e^{\frac{\beta\omega}{2}}]$.
The grand canonical pressure $P_L$ is  then  given from the grand
canonical
partition function $\Xi_L$ by

\begin{equation} \label{PL}
P_L(\beta,z,\omega)=\frac{1}{\beta L^3}\ln
\Xi_L(\beta,z,\omega)=\frac{1}{\beta L^3}{\rm Tr}\left [\ln\left
  (\mathbf{1} + z e^{-\beta H_L(\omega)}\right )\right ].
\end{equation}
Define for $\omega>0$,
$$P_{\infty}(\beta,z,\omega)=
\omega(2\pi\beta)^{-\frac{3}{2}}\,\sum_{k=0}^{\infty}
f_{\frac{3}{2}}(ze^{-(k+\frac{1}{2})\omega\beta})$$
and for $\omega=0$,
$$P_{\infty}(\beta,z,0)=\beta^{-1}(2\pi\beta)^{-\frac{3}{2}}\,f_{\frac{5}{2}}(z)$$
where $f_{\alpha}(z)$ are the standard Fermi functions (see e.g. \cite{AC}).
It  is  proved in \cite{AC} that
$P_L(\beta,z,\omega)$ admits
$P_{\infty}(\beta,z,\omega)$ as thermodynamic limit in the following
sense: for all $K$ compact included in $D$, and $\omega\geq 0$, one has
\begin{eqnarray}\label{lim-thermo-PL}
\lim_{L\rightarrow\infty}\,\sup_{z\in K} \left|
P_L(\beta,z,\omega)-P_{\infty}(\beta,z,\omega)\right| =0.
\end{eqnarray}
Yet it is known that if we define the grand canonical density
$\rho_L(\beta,z,\omega):=\beta z \frac{\partial P_L}{\partial
  z}(\beta,z,\omega)$ and
$\rho_{\infty}(\beta,z,\omega):=\beta z \frac{\partial
  P_{\infty}}{\partial z}(\beta,z,\omega)$,
one also has under same assumptions
\begin{eqnarray}\label{lim-thermo-rhoL}
\lim_{L\rightarrow\infty}\,\sup_{z\in K} \left|
\rho_L(\beta,z,\omega)-\rho_{\infty}(\beta,z,\omega)\right| =0.
\end{eqnarray}
The grand canonical generalized susceptibilities of a gas of fermions are
defined by:
\begin{eqnarray}\label{def-suscept-gen}
\chi_L^N(\beta,z,\omega):=\frac{\partial^N
  P_L}{\partial\omega^N}(\beta,z,\omega),\quad N\geq 1.
\end{eqnarray}

Notice that  $\chi_L^1(\beta,z,\omega)$ is the magnetization of the
system, and  $\chi_L^2(\beta,z,\omega)$ is the  magnetic
susceptibility. In case $\omega=0$ and $L\to\infty$, it is known in the
physical literature (see [A-B-N 1] for the rigorous proof) that
$\chi_\infty^1(\beta,z,0)=0$ and $\chi_\infty^2(\beta,z,0)\neq
0$. Thus at zero field, the magnetic response is quadratic. In case
$\omega\neq 0$, the magnetization is not zero, thus the magnetic
response becomes linear.
For $N\geq 2$, $\chi_L^N $ would give higher corrections to the linear
approximation.

Our main result is that the generalized
susceptibilities admit a thermodynamic limit in the following sense:
\begin{theorem}\label{mainthm}
Fix $N\in\N^*$, $\beta>0$ and $\omega\geq 0$. Then for every compact $K$
included in $D=\C\backslash\,]-\infty,-e^{\frac{\beta\omega}{2}}]$, one has
\begin{eqnarray}\label{lim-thermo-chiLn}
\lim_{L\rightarrow\infty}\,\sup_{z\in K} \left|
\chi_L^N(\beta,z,\omega)-\chi_{\infty}^N(\beta,z,\omega) \right| =0.
\end{eqnarray}
\end{theorem}
A straightforward consequence of this result is that under the same
assumptions:
\begin{eqnarray}
\forall N\in\N^*,\,\,\forall m\in\N,\,\,\lim_{L\rightarrow\infty}\,\sup_{z\in
K} \left| \frac{\partial^m\chi_L^N}{\partial
z^m}(\beta,z,\omega)-\frac{\partial^m\chi_{\infty}^N}{\partial
z^m}(\beta,z,\omega) \right| =0.
\end{eqnarray}
Having uniform limits with respect to $z$ is very useful if one wants
to translate this type of results in the canonical ensemble. See for
example [C 1] and [C 2] for further ideas in this direction.

Now let us mention some previous works dealing with similar
problems.  One  of  truly rigorous results for the case
$\omega=0$ and $N=1,2$ was given by Angelescu {\it et al} in [A-B-N
1]. Then Macris {\it et al} in [M-M-P] discussed the case when
$\omega$ was arbitrary but $N=1$ and $|z|<1$. In [C 1] this condition
on $z$ was lifted  for  the case of  Bose
statistics (but this  result can be immediately translated for the Fermi
case). Concluding, our present result settles the question for
derivatives of all order, for all Larmor frequencies and for all fugacities.

We would like to
remind the reader that this paper only
contains very basic ideas about proofs, and is mainly intended to give a
detailed overview about the long and rather complicated technical
steps that are needed.

%%%%%%%%%%%%%%%%%%%%%%%%%%%%%%%%%%%%%%%%
 \section{Strategy}%
%%%%%%%%%%%%%%%%%%%%%%%%%%%%%%%%%%%%%%%%
\vskip3mm\noindent Let us recall Vitali-Porter theorem \cite{H-P}:
%%%%%%%%%%%%%%%%%%%%%%%%%%%%%%%%%%%%%%%%

\begin{theorem}\label{vitali} Let $\{f_L\}_{L\geq 1}$ be a
  family of holomorphic functions on a fixed domain $D\subseteq \C$.
  Assume that $|f_L(z)|\leq M$ for all $L\geq 1$ and all $z\in D$. Assume
  also the existence of a subset $D'\subseteq D$ having an
  accumulation point $z_0\in D$, such that
  $\lim_{L\rightarrow\infty}f_L(z)$ exists for each $z\in D'$.
Then $\lim_{L\rightarrow\infty}f_L(z)$ exists everywhere in $D$, the
  convergence is uniform with respect to $z$ in any compact subset of
  $D$ and the limit function $f_\infty(z)$ is holomorphic in $D$.
\end{theorem}
In our case, we have $D=\C\setminus
\,]-\infty,-e^{\frac{\beta\omega}{2}}]$, $D'=\{z\in \C:\,|z|<1\}$,
$f_L(\cdot)=\chi_L^N(\beta,\cdot,\omega)$, and
$f_\infty(\cdot)=\chi_\infty^N(\beta,\cdot,\omega)$.

Therefore,  to prove the Theorem \ref{mainthm}, we need to  show  first the
pointwise
convergence on the unit open disk
\begin{eqnarray}\label{amontrerlemme3}
\lim_{L\rightarrow\infty}\chi_L^N(\beta,z,\omega)=\chi_{\infty}^N(\beta,z,\omega),\quad
|z|<1,
\end{eqnarray}
and second, that for every compact $K\subset D$  one has the uniform
bound in $L\geq 1$
\begin{eqnarray}\label{partie2}
\sup_{z\in K}\left|\chi_L^N(\beta,z,\omega)\right|\leq\,\cst(\beta,K,\omega).
\end{eqnarray}
Then Theorem \ref{mainthm} would be proven.
%%%%%%%%%%%%%%%%%%%%%%%%%%%%%%%%%%%%%%%%
\section{Elements of proofs}%
%%%%%%%%%%%%%%%%%%%%%%%%%%%%%%%%%%%%%%%%
\subsection{The pointwise limit: proof of (\ref{amontrerlemme3})}
%%%%%%%%%%%%%%%%%%%%%%%%%%%%%%%%%%%%%%%%
Denote by $({\cal I}_1(L^2(\Lambda)),\|\,\cdot \,\|_{{\cal I}_1})$ the
Banach space of trace class operators. It is well known that for any
$\omega\in\R$, the family of operators
$\{W_L(\beta,\omega)\}_{\beta>0}=\{e^{-\beta
  H_{L}(\omega)}\}_{\beta>0}$ is a Gibbs semigroup  [H-P] (the operators
$H_{L}(\omega)$
are self-adjoint, positive and $\{W_L(\beta,\omega)\}_{\beta>0} \subset
{\cal I}_1(L^2(\Lambda))$).
  On the other hand $W_L$
has an integral kernel $G_L(\x,\x',\beta,\omega)$ which is continuous on
$\Lambda\times\Lambda$ with respect to spatial variables $\x$ and
$\x'$. The diamagnetic inequality reads as:
\begin{equation}\label{diamagn}
|G_L(\x,\x',\beta,\omega)|\leq
 \frac{1}{(2\pi\beta)^{3/2}}\: e^{-\frac{|\x-\x'|^2}{2\beta}},\quad
 \x,\x'\in \Lambda,
\end{equation}
 which then implies that
\begin{eqnarray}\label{majI1WL}
\|W_L\|_{{\cal I}_1}=\tr\,W_L=\int_{\Lambda}\,
G_L(\x,\x,\beta,\omega)d\x\leq\frac{L^3}{(2\pi\beta)^{\frac{3}{2}}}
\end{eqnarray}
 Suppose $z$ in the unit disk $D'$,  from (\ref{PL}) we  have  (see e.g.
(2.10) and (2.11) in [C 1]
in the case of Fermi statistics):
\begin{eqnarray}\label{def-PL-disque-unite}
P_L(\beta,z,\omega)=\frac{1}{\beta L^3}
\sum_{n=1}^{\infty}\frac{(-1)^{n+1}z^n}{n}\,{\rm
Tr}\left(W_L\left(n\beta,\omega\right)\right).
\end{eqnarray}
We denote by $G_{\infty}(\x,\x',\beta,\omega)$ the integral kernel of
the corresponding operator defined on the whole space: $\{e^{-\beta
  H(\omega)}\}_{\beta>0} $ (see (2.2) in
[A-C] or (4.90) in [C 1]). Its diagonal is very  simple and
 is given by
$$G_{\infty}(\x,\x,\beta,\omega)=\frac{1}{(2\pi\beta)^{3/2}}\frac{\omega\beta/2}{\sinh(\omega\beta/2)}.$$
We remark that this quantity is $\x-$independent  and  in view of (\ref{PL})
we can
write
\begin{eqnarray}\label{defPinfini-disque-unite}
P_{\infty}(\beta,z,\omega)=\frac{1}{\beta}
\sum_{n=1}^{\infty}\frac{(-1)^{n+1}z^n}{n}\,\,G_{\infty}(\x,\x,n\beta,\omega).
\end{eqnarray}

We are interested in derivatives of $P_L$ with respect to
$\omega$. Due to formula (\ref{def-PL-disque-unite}), these
derivatives will act on the trace of the semigroup. We are thus
motivated to study the ${\cal I}_1$-analyticity with respect
to $\omega$ of the semigroup. Although this result was already proven
in [A-B-N 1], we state it in Lemma
\ref{anai1}, since we will use it later on.

In order to do that, we need to introduce further notation.
Define the following  operators by their corresponding integral kernels:
\begin{eqnarray}\nonumber
\hat{R}_{1,L}(\x,\x',\beta,\omega) &:=&  \A(\x)\cdot(i\nabla_\x+\omega
\A(\x))G_L(\x,\x',\beta,\omega),\\  \label{def-R1L-R2L}
\hat{R}_{2,L}(\x,\x',\beta,\omega) &:=&
\frac{1}{2}\;\A^2(\x)\,G_L(\x,\x',\beta,\omega).
\end{eqnarray}
The  operators  $\hat{R}_{1,L}(\beta,\omega)$ and
$\hat{R}_{2,L}(\beta,\omega)$ are of trace class  as well as the operators,

\begin{eqnarray}\label{defInL}
&{}&
I_{n,L}(i_1,...,i_n)(\beta,\omega):=\int_0^{\beta}d\tau_1\int_0^{\tau_1}d\tau_2...\int_0^{\tau_{n-1}}d\tau_n\,W_L(\beta-\tau_1,\omega)\\
&{}&  \cdot
\hat{R}_{i_1,L}(\tau_1-\tau_2,\omega)\hat{R}_{i_2,L}(\tau_2-\tau_3,\omega)...\hat{R}_{i_{n-1},L}(\tau_{n-1}-\tau_n,\omega)\hat{R}_{i_n,L}(\tau_n,\omega)\nonumber
\end{eqnarray}
for $n\geq 1$ and for $(i_1,...,i_n)\in\{1,2\}^{n}$. We can finally give the analyticity
result, define
\begin{eqnarray}\label{fctcar}
c_n^{N}(i_1,...,i_n):=\left\{\begin{array}{ccc} 1 &&{\rm if}\,\,
    i_1+...+i_n=N \\ 0 && {\rm  otherwise .}\end{array}\right.
\end{eqnarray}
%%%%%%%%%%%%%%%%%%%%%%%%%%%%%%%%%
\begin{lemma}\label{anai1}
%%%%%%%%%%%%%%%%%%%%%%%%%%%%%%%%
The operator-valued function $\R\ni \omega\mapsto W_L(\beta,\omega)\in
{\cal I}_1$
admits an entire extension to $\C$.  Fix $\omega_0\in\R$. For all
$\omega\in\C$ we have
\begin{eqnarray}\label{deriveeWL}
W_L(\beta,\omega) &=&
  \sum_{N=0}^{\infty}\frac{(\omega-\omega_0)^N}{N!}\, \, \frac{\partial^N
  W_L}{\partial \omega^N}(\beta,\omega_0), \\
\frac{\partial^N
  W_L}{\partial \omega^N}(\beta,\omega_0) &=&
  N!\sum_{n=1}^{N}(-1)^n\sum_{i_j\in\{1,2\}}
c_n^N(i_1,...,i_n) I_{n,L}(i_1,...,i_n)(\beta,\omega_0).\nonumber
\end{eqnarray}
\end{lemma}
This implies in particular that the traces of the semigroup $W_L$ which
appear in formula (\ref{def-PL-disque-unite}) are entire functions
of $\omega$.

\begin{remark}\label{rem1}
It is important to notice that the  expansion  (\ref{deriveeWL}) is not
really convenient
if one wants to prove (\ref{amontrerlemme3}). That is because  the
expressions  (\ref{def-R1L-R2L})  contain  at least one term as  $\A(\x)$
which
behaves like $\x$.  So direct estimates show that
the trace norm of $\frac{\partial^N
  W_L}{\partial\omega^N}(\beta,\omega_0)$ (see e.g. [A-B-N 2]) behaves like
$L^{3+N}$, and
this is very far from the desired behavior of $L^3$. It is true that
when we look at the trace and not at the trace-norm, things are quite
different. In [A-B-N 1] it is proved at $\omega_0=0$ and for $N=1,2$ that due
to some
remarkable identities, the terms growing like $L^{3+N}$ are
identically zero. What we do next in our paper is to give an
alternative expansion which takes care of these singularities for all
terms at the same time.
\end{remark}
In order to do that, we concentrate on the kernel
$G_L$. We remark first the following
%%%%%%%%%%%%%%%%%%%%%%%%%%%%%%%%%
\begin{lemma}\label{lemma2}
%%%%%%%%%%%%%%%%%%%%%%%%%%%%%%%%
For every $\omega\in \C$, the operator $W_L(\beta,\omega)$ defined by
the series in (\ref{deriveeWL}) admits an
integral kernel $G_L(\x,\x',\beta,\omega)$. This kernel is defined as
the sum of a series  as  in (\ref{deriveeWL}) where instead of
operators we consider their integral kernels. Then $G_L$ is continuous
with respect to the spatial variables, and is an entire function of
$\omega$. In addition, for all $\x,\x'\in\Lambda$ fixed, one has
\begin{eqnarray}\label{derivnoyau=noyauderiv}
\frac{\partial^N G_L}{\partial\omega^N}\,(\x,\x',\beta,\omega)=
\left(\frac{\partial^N W_L}{\partial\omega^N}\right)\,(\x,\x',\beta,\omega).
\end{eqnarray}
\end{lemma}
\noindent {\bf Hints to the proof}.
Notice that $\frac{\partial^N G_L}{\partial\omega^N}(\x,\x')$ is kernel's
derivative, while $\frac{\partial^N W_L}{\partial\omega^N}(\x,\x')$
denotes the kernel of the trace class operator $\frac{\partial^N
  W_L}{\partial\omega^N}$. The idea of the proof consists in showing
that when replacing $I_{n,L}$ from (\ref{deriveeWL}) with its integral
kernel (defined as a continuous function in $\x,\x'\in
\overline{\Lambda}$ by the multiple convolution (\ref{defInL})), the
power series in (\ref{deriveeWL}) converges uniformly for $\x,\x'\in
\overline{\Lambda}$, and has an infinite radius of
convergence. The estimates rely on the diamagnetic inequality
(\ref{diamagn}) and an
induction argument. \qed

\vspace{0.5cm}

 Since $\frac{\partial^N G_L}{\partial\omega^N}(\x,\x',\beta,\omega)$ is
 continuous for
$(\x,\x')\in\Lambda\times\Lambda$, and $\frac{\partial^N
 W_L}{\partial\omega^N}$ is a trace class operator, its trace can be
expressed as the integral of the
 diagonal of its kernel (see the remark at page 523 in [K]).
We conclude that for every $\omega\in\C$:
\begin{eqnarray}\label{tr-WL-deriv}
\frac{\partial^N}{\partial\omega^N}{\rm
Tr}\left(W_L\left(\beta,\omega\right)\right)={\rm Tr}\left(\frac{\partial^N
W_L}{\partial\omega^N}\left(\beta,\omega\right)\right)=\int_{\Lambda}\frac{\partial^N
G_L}{\partial\omega^N}\,(\x,\x,\beta,\omega)\,d\x.
\end{eqnarray}

In the light of Remark \ref{rem1}, we need a different formula for the
above kernel, so that the apparent growing terms cancel each
other. This will be done by using a modified perturbation theory for
magnetic Gibbs semigroups. Previous works which dealt with similar
problems are [C-N], [C 1], [B-C] and [N].

For that, we introduce the magnetic phase $\phi$ and the magnetic flux $\fl$
where for $\x,\y,\z\in\Lambda$,
and as before ${\bf e}_3=(0,0,1)$,
\begin{equation}\label{phaseflux}
 \phi(\x,\y):=\frac{1}{2}\,{\bf e}_3\cdot (\y\wedge
\x),\,\,\,\,\fl(\x,\y,\z):=\phi(\x,\y)+\phi(\y,\z)+
\phi(\z,\x).
\end{equation}
We have
\begin{equation} \label{estfl}
|\fl(\x,\y,\z)|\leq\,|\x-\y|\, |\y-\z|.
\end{equation}
For every $n\geq 1$ and for every points
$\x, \y_1,..., \y_n\in\Lambda$, we introduce
$${\Fl}_1(\x,\y_1)=0,\quad {\Fl}_n(\x,\y_1,...,\y_n)= \sum_{k=1}^{n-1}\fl
(\x,\y_k,\y_{k+1}),\; n\geq 2.$$
Fix $\omega >0$. Consider now the bounded operators given by their integral
kernels:
\begin{eqnarray}\label{def_R1L-R2L-RL}
R_{1,L}(\x,\x',\beta,\omega)& := & \A(\x-\x')\cdot
\left(i\nabla_\x+\omega
  \A\left(\x\right)\right)G_L(\x,\x',\beta,\omega),\nonumber \\
 R_{2,L}(\x,\x',\beta,\omega)& := & \frac{1}{2}\,
\A^2(\x-\x')G_L(\x,\x',\beta,\omega)
\end{eqnarray}
and for all $\x\in\Lambda$, $n\geq 1$ and $k\geq 0$
\begin{eqnarray}\nonumber
&& \!\!\!\!\!\!\!\!\!\!\!\!\!\!\!\!\!\!
W_{L,n}^k(\x,\x,\beta,\omega):=\sum_{j=1}^n(-1)^j\sum_{(i_1,...,i_j)\in\{1,2\}^{j}}c_{j}^{n}(i_1,...,i_j)
\int_0^{\beta}d\tau_1\int_0^{\tau_1}d\tau_2\,...\\ \nonumber
&& \!\!\!\!\!\!\!\!\!\!\!\!\!\!\!\!\!\!
\int_0^{\tau_{j-1}}d\tau_j\int_{\Lambda}d\y_1\,...\int_{\Lambda}d\y_j\,\frac{\left(i\left(\Fl_{j}(\x,\y_1,...,\y_{j})\right)\right)^k}{k!}\,\,G_L(\x,\y_1,\beta-\tau_1,\omega)\\
\label{def-WL-n-k}
&& \!\!\!\!\!\!\!\!\!\!\!\!\!\!\!\!\!\!
R_{i_1,L}(\y_1,\y_2,\tau_1-\tau_2,\omega)\,...\,R_{i_{j-1},L}
(\y_{j-1},\y_j,\tau_{j-1}-\tau_j,\omega)R_{i_j,L}(\y_j,\x,\tau_j,\omega).
\end{eqnarray}
 By convention, in the case when
$k=0$ we set $0^0\equiv 1$.

The next lemma gives a new expression for the diagonal of kernel's
$N$-th derivative with respect to $\omega$ at finite volume.
%%%%%%%%%%%%%%%%%%%%%%%%%%%%%%%%%
\begin{lemma}\label{lemmaa3}
%%%%%%%%%%%%%%%%%%%%%%%%%%%%%%%%
Fix $\omega_0\geq 0$. Then for all $\x\in\Lambda$, and for all $N\in\N^*$,
one has
\begin{eqnarray}\label{derivee-noyau-diag-formule-2}
\frac{1}{N!}\,\frac{\partial^N G_L}{\partial\omega^N}(\x,\x,\beta,\omega_0)=\sum_{n=1}^N
W_{L,n}^{N-n}(\x,\x,\beta,\omega_0).
\end{eqnarray}
\end{lemma}

\noindent {\bf Hints to the proof}. The proof heavily relies on the
general theory developed in [A-B-N 2] and in [C 1], where a version of Duhamel's
formula
is written for the perturbed semigroup $W_L(\beta,\omega)$ for
$d\omega:=\omega-\omega_0$ small, here $\omega_0\geq 0$ is fixed  (see
Proposition 3 and formula
(4.61) in [C 1]). Roughly speaking, one has to iterate this formula
$N$ times, and identify the term containing $(d\omega)^N$.\qed

\vspace{0.5cm}

%%%%%%%%%%%%%%%%%%%%%%%%%%%%%%%%%%
Based on the above formula, we can give an expression for the
corresponding quantities at infinite volume:
%%%%%%%%%%%%%%%%%%%%%%%%%%%%%%%%%
\begin{lemma}\label{lemaa4}
%%%%%%%%%%%%%%%%%%%%%%%%%%%%%%%%
Fix $\omega_0\geq 0$. Then for all $x\in\Lambda$, and for all $N\in\N^*$, one
has
\begin{eqnarray}\nonumber
&& \!\!\!\!\!\!\!\!\! \frac{1}{N!}\frac{\partial^N
  G_{\infty}}{\partial\omega^N}\,(\x,\x,\beta,\omega_0)=
\lim_{L\rightarrow\infty}\,\frac{1}{N!}\frac{\partial^{N}G_L}{\partial\omega^{N}}(\x,\x,\beta,\omega_0)\\
\nonumber
&& \!\!\!\!\!\!\!\!\!
=\sum_{n=1}^{N}\sum_{j=1}^{n}(-1)^{j}\sum_{i_k\in\{1,2\}}
c_{j}^{n}(i_1,...,i_j)\int_0^{\beta}d\tau_1\int_0^{\tau_1}d\tau_2...\int_0^{\tau_{j-1}}\!\!d\tau_j\int_{\R^3}d\y_1...\\
\nonumber
&& \!\!\!\!\!\!\!\!\!\int_{\R^3} d\y_j\,\frac{(i
\Fl_{j}(\x,\y_1,...,\y_{j}))^{N-n}}{(N-n)!}\,\,G_{\infty}(\x,\y_1,\beta-\tau_1,\omega_0)R_{i_1,\infty}(\y_1,\y_2,\tau_1-\tau_2,\omega_0)\\
\label{lim-des-noyaux-diag}
&& \!\!\!\!\!\!\!\!\!
  ...R_{i_{j-1},\infty}(\y_{j-1},\y_j,\tau_{j-1}-\tau_j,\omega_0)
  R_{i_j,\infty}(\y_j,\x,\tau_j,
\omega_0)
\end{eqnarray}
where $R_{j,\infty}$ are defined as in (\ref{def_R1L-R2L-RL}) with
$L\equiv \infty$.
\end{lemma}

\noindent {\bf Hints to the proof}. We need to estimate
$(G_L-G_{\infty})(\x,\x')$ and $(i\nabla_\x+\omega_0
\A(\x))(G_L-G_{\infty})(\x,\x')$. We have to take into account the
walls' influence on the integral kernel at finite
volume. We use a variant of Green's formula for the solutions of the
heat equation inside $\Lambda$. Thus, we  get  the next result.
Define for all  $\x\in\Lambda$,
$\delta_\x:={\rm dist}(\x,\partial\Lambda)$ and $M:=\{\x\in\Lambda\;
:\,\delta_\x\leq 1\}$. Then one has
\begin{eqnarray}\nonumber
&& \!\!\!\!\!\!\!\!\!\!\!\!\!\!\!\!\!\!
|(G_L-G_{\infty})(\x,\x',\beta,\omega_0)|,\,
|(i\nabla_\x+\omega_0 \A(\x))(G_L-G_{\infty})(\x,\x',\beta,\omega_0)|\\
\label{influencebord}
&& \!\!\!\!\!\!\!\!\!\!\!\!\!\!\!\!\!\!
\leq\,\frac{\cst}{\beta^2}\,\,{\rm
exp}\left(-\frac{|\x-\y|^{2}}{c\beta}\right)\left(\chi_{M}(\x)+\chi_{M}(\x')+{\rm
exp}\left(-\frac{\delta_{\x}^{2}}{c\beta}-
\frac{\delta_{\x'}^{2}}{c\beta}\right)\right),
\end{eqnarray}
where $c>0$ is a constant and $\chi_M$ is the
characteristic function of $M$. Then (\ref{lim-des-noyaux-diag})
follows after some straightforward calculations from
(\ref{diamagn}), (\ref{influencebord}), (\ref{phaseflux}), the
estimate (4.46) in [C 1], and (\ref{derivee-noyau-diag-formule-2}). \qed

%%%%%%%%%%%%%%%%%%%%%%%%%%%%%%%%%%%%%%%%%%%%%%%%%%%%%%%%%%%%%%%%%%%%%%%%%%%%%%%%%%%%%%%%%%%%%%%%%%%%%%%%%%%%%

\vspace{0.5cm}

Since our aim is to prove (\ref{amontrerlemme3}), we write from
(\ref{def-PL-disque-unite}) and
(\ref{defPinfini-disque-unite})
\begin{eqnarray}\label{PL-Pinfini}
(P_L-P_{\infty})(\beta,z,\omega)=\frac{1}{\beta
L^3}\sum_{n=1}^{\infty}\frac{(-1)^{n+1}
z^n}{n}\,\int_{\Lambda}d\x\,(G_L-G_{\infty})(\x,\x,n\beta,\omega).
\end{eqnarray}

In order to conclude that (\ref{amontrerlemme3}) is true, it will be
sufficient to
show that
%%%%%%%%%%%%%%%%%%%%%%%%%%%%%%%%%
\begin{lemma}\label{lemma6}
%%%%%%%%%%%%%%%%%%%%%%%%%%%%%%%%
Fix $\omega_0\geq 0$. Then for all $N\in\N^* $, one has
%%%%%%%%%%%%%%%%%%%%%%%%%%%%%%%%
\begin{eqnarray}\label{majo-L2}
\left|\int_{\Lambda}d\x\,\left(\frac{\partial^N
G_L}{\partial\omega^N}(\x,\x,\beta,\omega_0)-\frac{\partial^N
G_{\infty}}{\partial\omega^N}(\x,\x,\beta,\omega_0)\right)\right|\leq\,L^2\,f(\beta,\omega_0,N),
\end{eqnarray}
where $f(\cdot,\omega_0,N)$ is a function of $\beta$ which is
polynomially bounded.
\end{lemma}

\noindent {\bf Hints to the proof}. Fix $N\in\N^*$. One denotes
by $F_{L,N}(\x,\beta,\omega_0)$ the formula obtained by replacing in formula
(\ref{lim-des-noyaux-diag}) all the spatial integrals on $\R^3$ by integrals
on $\Lambda$. To estimate the difference
$F_{L,N}(\x,\beta,\omega_0)-\frac{\partial^N
  G_{L}}{\partial\omega^N}(\x,\x,\beta,\omega_0)$, we use
(\ref{derivee-noyau-diag-formule-2}),
(\ref{diamagn}), (\ref{influencebord}), (\ref{phaseflux}), and the estimate
(4.46) in [C 1].
Finally we find that
\begin{eqnarray}\label{ddd333}
\left|\int_{\Lambda}d\x\,\left(F_{L,N}(\x,\beta,\omega_0)-\frac{\partial^N
  G_{L}}{\partial\omega^N}(\x,\x,\beta,\omega_0)\right)\right|\leq
L^2\,f_1(\beta,\omega_0,N),
\end{eqnarray}
where $f_1$ is polynomially bounded with respect to $\beta$.
Now we need to estimate the difference
$F_{L,N}(\x,\beta,\omega_0)-\frac{\partial^N
  G_{\infty}}{\partial\omega^N}(\x,\x,\beta,\omega_0)$. From the
  definition of $F_{L,N}$, this difference will consist with integrals
  as in (\ref{lim-des-noyaux-diag}) where in at least one of the
  spatial integrals one integrates over $\R^3\setminus \Lambda$. Since
  it can also be shown that (\ref{ddd333}) holds again if we replace
  $G_L$ with $G_\infty$, the lemma is proven up to the use of the
  triangle inequality which yields (\ref{majo-L2}).  \qed

\vspace{0.5cm}

Since we have seen in (\ref{tr-WL-deriv}) that the derivatives with
respect to $\omega$ and the trace
commute at finite volume, formula (\ref{PL-Pinfini}) and Lemma \ref{lemma6}
show that for every $|z|<1$, the derivatives
with respect to $\omega$ of $(P_L-P_{\infty})(\beta,z,\omega)$ behave
like $\frac{1}{L}$, which finishes the proof of (\ref{amontrerlemme3}).
%%%%%%%%%%%%%%%%%%%%%%%%%%%%%%%%%%%%%%%%
\subsection{The uniform bound: proof of (\ref{partie2})}
%%%%%%%%%%%%%%%%%%%%%%%%%%%%%%%%%%%%%%%%
If we denote by
$g_L(\beta,\tau,\omega)=g_L(\beta,\tau,\omega,z,\xi)=(\xi-zW_L(\beta,\omega))^{-1}zW_L(\tau,\omega)$
then for $\beta>0$, $z\in K\subset
D$, $\omega\geq 0$,
one has (see (4.2) in [C 1] for the Bose case):
\begin{eqnarray}\label{PL-gL}
P_L(\beta,z,\omega)=\frac{1}{2i\pi}\int_{C}
d\xi\,\,\frac{\ln(1+\xi)}{\xi}\,{\frac{1}{\beta L^3}}\,{\rm
Tr}\left(g_L\left(\beta,\beta,\omega\right)\right)
\end{eqnarray}
with $C\subset\C\backslash ]-\infty,-1]$ surrounds the eigenvalues of
$zW_L$. In addition, one chooses $C$ in order to have
$(\xi-zW_L(\beta,\omega))^{-1}$ bounded for every $\xi\in C$ and $z\in
K$. We have seen that $\omega\mapsto W_L(\beta,\omega)$ is ${\cal
  I}_1$-analytic on $\C$; one can also see that $\tr(g_L)$ is a
real-analytic function of $\omega$. In the end, the generalized
susceptibilities are well defined as functions of $z$ on $D$. We see that
(\ref{partie2}) will follow from
\begin{eqnarray}\label{amontrer2}
\sup_{z\in K}\sup_{\xi\in {\cal
C}}\left|\frac{\partial^N}{\partial\omega^N}\,\tr\,(g_L(\beta,\beta,\omega))\right|\leq\,L^3\,\cst(\beta,K).
\end{eqnarray}
Here we have a similar problem as the one pointed out for the
semigroup in Remark \ref{rem1}. We could try to use the inequality

$$\left|\frac{\partial^N}{\partial\omega^N}\,\tr\,(g_L(\beta,\beta,\omega))\right
|\leq
\left
  \Vert\frac{\partial^N g_L}{\partial\omega^N}(\beta,\beta,\omega)\right
\Vert _{{\cal
    I}_1},$$
but the right hand side behaves like $L^{3+N}$ and not like $L^3$ as
desired. What we do instead is finding a Taylor expansion {\it directly for
  the trace}, and to give the right estimate for its derivatives.

In view of developing $\tr(g_L)$ as a function of $\omega$ in a small
real neighborhood $\Omega$ of $\omega_0\geq 0$, we first analyze
$g_L$ in ${\cal I}_1(L^2(\Lambda))$. As a general rule, for an
integral operator $T(\omega_0)$ with kernel $t(\x,\x',\omega_0)$, we
denote by $\tilde{T}(\omega)$ the operator which has an integral
kernel given by (see also (\ref{phaseflux}))
$$\tilde
{t}(\x,\x',\omega):=e^{i(\omega-\omega_0)\phi(\x,\x')}t(\x,\x',\omega_0).$$

%%%%%%%%%%%%%%%%%%%%%%%%%%%%%%%%%
\begin{lemma}\label{lemma7}
%%%%%%%%%%%%%%%%%%%%%%%%%%%%%%%%
Fix $\omega_0\geq 0$ and $N\geq 1$.
Then for every $d\omega:=\omega-\omega_0$ small, there exist $N$ trace
class operators
$a_{L,n}(\beta,\omega)=a_{L,n}(\beta,\omega,z,\xi)$,
$1\leq n\leq N$, and
an operator
$R_{L,N+1}(\beta,\omega)=R_{L,N+1}(\beta,\omega,z,\xi)$
such that
\begin{eqnarray}\label{dev-I1-gL}
g_L(\beta,\beta,\omega) &=&
\widetilde{g_L}\left(\beta,\frac{\beta}{2},\omega\right)\widetilde{W_L}\left(\frac{\beta}{2},\omega\right)+\sum_{n=1}^{N}
d\omega^{n}\,a_{L,n}(\beta,\omega)\\  \nonumber
&+& R_{L,N+1}(\beta,\omega),\end{eqnarray}
where
\begin{eqnarray}\label{dev-I1-gL2}
\|a_{L,n}(\beta,\omega)\|_{{\cal
    I}_1} &\leq &\cst(\beta,K,C)L^3,\quad \omega\in \Omega, \quad
    1\leq n\leq N \nonumber \\
\|R_{L,N+1}(\beta,\omega)\|_{{\cal
    I}_1} &\leq & |d\omega|^{N+1}\,\cst(\beta,K,C) L^3.
\end{eqnarray}
Notice that the operators $a_{L,n}$ still depend on $\omega$.
\end{lemma}

\noindent {\bf Hints to the proof}. We use a technique which
generalizes the one developed in [C 1], which only worked for
$N=1$. Our generalization is considerably more involved than the
original argument given in [C 1], which at its turn was rather
lengthy. See for comparison formula (4.84) in [C 1], which corresponds
to the case $N=1$ in our lemma. Full proofs will be given elsewhere. \qed

\vspace{0.5cm}

%%%%%%%%%%%%%%%%%%%%%%%%%%%
As (\ref{dev-I1-gL}) is valid in ${\cal I}_1(L^2(\Lambda))$, we can
take the trace term by term in this equality.  This gives (see also (4.85)
in [C 1]):
\begin{eqnarray}\nonumber
\tr(g_L(\beta,\beta,\omega)) &=& \tr(g_L(\beta,\beta,\omega_0))\\
\label{dev-tr}
&+& \sum_{n=1}^N
d\omega^n\left(\tr(a_{L,n}(\beta,\omega))\right)+\tr\left(R_{L,N+1}(\beta,\omega)\right).
\end{eqnarray}
The last technical result we need is contained in the following lemma,
given again without proof:

%%%%%%%%%%%%%%%%%%%%%%%%%%%%%%%%%
\begin{lemma}\label{lemma8}
%%%%%%%%%%%%%%%%%%%%%%%%%%%%%%%%
Fix $\omega_0\geq 0$, $N\geq 1$ and $1\leq n\leq N$. Then there exists a
family of
$\omega$-independent coefficients
$\{b_{L,n}^m(\beta,\omega_0)\}_{m\in\N}=\{b_{L,n}^m(\beta,\omega_0,z,\xi)\}_{m\in\N}$, and a remainder
$r_{L,n}^{N+1}(\beta,\omega)=r_{L,n}^{N+1}(\beta,\omega,z,\xi)$
such that for $d\omega=\omega-\omega_0$ small, one has
\begin{eqnarray}\label{bLnm}
\tr\left(a_{L,n}(\beta,\omega)\right)&=&
\sum_{m=0}^{N}d\omega^m\,b_{L,n}^m(\beta,\omega_0)+r_{L,n}^{N+1}(\beta,\omega),\\
\nonumber
|r_{L,n}^{N+1}(\beta,\omega)|&\leq& L^3 |d\omega|^{N+1}\,\cst(\beta,
K,C),\nonumber \\
\left|b_{L,n}^m(\beta,\omega_0)\right| &\leq & L^3\,\cst(\beta,K,C).\nonumber
\end{eqnarray}
\end{lemma}

\vspace{0.5cm}

%%%%%%%%%%%%%%%%%%%%%%%%%%%
\vskip3mm\noindent Consequently, from (\ref{dev-tr}) and (\ref{bLnm}) we have

\begin{eqnarray}
\left [\frac{\partial^N}{\partial\omega^N}\tr(g_L)\right
](\beta,\beta,\omega_0,z,\xi)=\sum_{n=1}^{N}b_{L,n}^{N-n}
(\beta,\omega_0,z,\xi),
\end{eqnarray}
and (\ref{amontrer2}) follows.

\vspace{0.5cm}

\noindent {\bf Acknowledgments}. Two of the authors (H.C. and D.L.) wish to
thank the French Embassy
in Copenhagen for financial support  for a visit 
 to Centre de Physique Th\' eorique-CNRS-Marseille (H.C.) and
for a visit   to Department of Mathematical Sciences-Aalborg  (D.L.).
%%%%%%%%%%%%%%%%%%%%%%%%%%%%%%%%%%%%%%%%%%%%%%%%%%%


\begin{thebibliography}{2}
%%%%%%%%%%%%%%%%%%%%%%%%%%
\bibitem[A-B-N 1]{ABN1} Angelescu, N., Nenciu, G., Bundaru, M.: On the
  Landau Diamagnetism. Commun. in Math. Phys., {\bf 42}, 9-28 (1975)
\bibitem[A-B-N 2]{ABN2} Angelescu, N., Nenciu, G., Bundaru, M.: On the
  perturbation of Gibbs semigroups. Commun. in Math. Phys., {\bf 42}, 29-30
(1975)
\bibitem[A-C]{AC} Angelescu, N., Corciovei, A..: On free quantum gases
  in a homogeneous magnetic field.
Rev. Roum. Phys., {\bf 20}, 661-671 (1975)
\bibitem[B-C]{Cb} Briet, P., Cornean, H.D.: Locating the spectrum for
  magnetic Schr\"odinger and Dirac operators.
Comm. Partial
  Differential Equations {\bf 27}, no. 5-6, 1079--1101 (2002)
\bibitem[C 1]{C} Cornean, H.D.: On the magnetization of a charged Bose
  gas in the ca\-no\-ni\-cal ensemble. Commun. in Math. Phys., {\bf
  212}, 1-27 (2000).
\bibitem[C 2]{C2} Cornean, H.D.: Magnetic response in ideal quantum
  gases: the thermodynamic limit. To appear in Markov Procces.
  Related Fields.
\bibitem[C-N]{Cn} Cornean, H.D., Nenciu, G.: On the eigenfunction decay
  for two dimensional magnetic Schr\"odinger operators. Commun. in Math.
Phys., {\bf
  192}, 671-685 (1998).
\bibitem[H-P]{H-P}Hille, E., Phillips,R.: Functional Analysis and
semi-groups. Providence, Rhode Island: American Mathematical Society, 1957
\bibitem[K]{K} Kato, T.: Perturbation Theory for Linear
  Operators. New York: Springer-Verlag, 1966
\bibitem[M-M-P]{ad} Macris, N., Martin, Ph.A., Pul\'e, J.V.:
Large volume asymptotic of Brownian integrals and orbital magnetism.
Ann. I.H.P. Phys. Theor. {\bf 66}, 147-183 (1997)
\bibitem[N]{adn} Nenciu, G.: On asymptotic perturbation theory for
  quantum mechanics:
almost invariant subspaces and gauge invariant magnetic perturbation
  theory. J. Math. Phys.,
 {\bf 43}, no. 3, 1273--1298 (2002)
\end{thebibliography}
\end{document}